\documentclass[12pt,preprint]{aastex}
\usepackage{times,color}
\def\mbf#1{\mbox{\boldmath ${#1}$}}

\shorttitle{Structure and Stability of Phase Transition Layers}
\shortauthors{Inoue et al.}
\begin{document}
\title{Structure and Stability of Phase Transition Layers in The Interstellar Medium}
\author{
Tsuyoshi Inoue\altaffilmark{1},
Shu-ichiro Inutsuka\altaffilmark{1} and
Hiroshi Koyama\altaffilmark{2}
}
\affil{$^1$Department of Physics, Graduate Scool of Science, Kyoto University, Sakyo-ku, Kyoto 606-8588, Japan}
\email{tsuyoshi@tap.scphys.kyoto-u.ac.jp, inutsuka@tap.scphys.kyoto-u.ac.jp}
\affil{$^2$Department of Earth and Planetary System Science, Graduate School of Science and Technology, Kobe University, Nada, Kobe 657-8501, Japan}
\email{hkoyama@kobe-u.ac.jp}
\begin{abstract}
We analyze the structure and stability of the transition layer (or front) that connects the cold neutral medium and warm neutral medium in the plane-parallel geometry.
Such fronts appear in recent numerical simulations of a thermally bistable interstellar medium.
The front becomes an evaporation or condensation front depending on the surrounding pressure.
The stability analysis is performed in both long- and short-wavelength approximations.
We find that the plane-parallel evaporation front is unstable under corrugational deformations, whereas the condensation front seems to be stable.
The instability is analogous to the Darrieus-Landau instability in combustion front.
The growth rate of the instability is proportional to the speed of the evaporation flow and the corrugation wavenumber for modes with wavelength much longer than the thickness of the front, and it is suppressed at scales approximately equal to the thickness of the front.
The timescale of the instability is smaller than the cooling timescale of the warm neutral medium ($\sim 1$ Myr), and can be as small as the cooling timescale of the cold neutral medium ($\sim 0.01-0.1$ Myr).
Thus, this instability should be one of the processes for driving the interstellar turbulence.
\end{abstract}
\section{INTRODUCTION}
It is widely known that the low- and mid-temperature parts of the interstellar medium (ISM) are composed of a thermally bistable fluid that results from the balance of radiative cooling and heating due to external radiation fields and cosmic rays (Field, Goldsmith \& Habing 1969; Wolfire et al. 1995, 2003).
The two stable phases are called the cold neutral medium (CNM) with temperature $T\approx 10-10^{2}$ K and the warm neutral medium (WNM) with temperature $T\approx 10^{4}$ K.
Zel'dovich \& Pikel'ner (1969) and Penston \& Brown (1970) studied the transition layer or front that connects these stable phases assuming plane-parallel geometry.
They found that a static solution is obtained at the so-called the saturation pressure, and steady phase exchange (evaporation and condensation) solutions are obtained when the pressure deviates from this value.

From an observational point of view, interstellar clouds are characterized by suprathermal emission-line widths, which can be related to the supersonic internal motion of gas, or ``turbulence," in clouds.
Recent numerical simulations have shown that supersonic turbulence in an isothermal or an adiabatic medium decays quickly, that is, within a crossing time, irrespective of the effect of magnetic fields (Mac Low et al. 1998; Stone et al. 1998; Ostriker et al. 1999).
In contrast, Koyama \& Inutsuka (2002) found long-lasting turbulent motion of tiny cloudlets in the WNM in a numerical simulation of the two-phase medium, which are formed as a consequence of thermal instability (see also Koyama \& Inutsuka 2004, Inutsuka \& Koyama 2004).
Audit \& Hennebelle (2005) found similar energetic turbulent motions in colliding flows, a topic has been studied further by Heitsch et al. (2005) and Vazquez-Semadeni et al. (2006).
However, a detailed analysis of the mechanism that maintains such turbulent motions remains to be done.

The two-phase evaporation front is analogous to a combustion front.
The latter are unstable to corrugational deformations, a phenomenon known as the Darrieus-Landau instability (DLI; see, e.g., Landau \& Lifshitz 1987; Zel'dovich et al. 1985).
From experiments and numerical simulations of chemical and nuclear flames, the DLI is known to develop into turbulence in the nonlinear regime, and this turbulence leads to accelerated combustion (Sivashinsky 1983; Niemeyer \& Hillebrandt 1995; Blinnikov et al. 1995).
It is also known that there is a similar instability that applies to evaporation fronts under the discontinuous-front approximation (Aranson et al. 1995; Inutsuka et al. 2005).
Thus, the instability is expected to be a driving mechanism of turbulence in the two-phase medium.

However, in the framework of the discontinuous front approximation, we cannot determine the most unstable scale of the instability and its growth timescale owing to the neglect of the thickness of the front.
Therefore, in this paper we undertake a linear stability analysis of the two-phase front by accounting for the effect of the finite thickness of the front.

The paper is organized as follows.
The steady smooth fronts that describe the structure of the transition layers are calculated numerically in \S 2.
In \S 3, we briefly review the linear stability of the front under corrugational deformations in the framework of the discontinuous-front approximation, which corresponds to analysis in the long-wavelength limit.
Then the linear stability of the smooth transition layers that are obtained in \S 2 is studied under the isobaric approximation which gives an appropriate result for small wavelength modes.
Finally, we summarize our results and discuss their implications in \S 4. 

\section{STRUCTURE OF STEADY FRONTS}

We start the with fluid equations describing the dynamics of an unmagnetized, optically thin ideal gas of density $\rho$, temperature $T$, pressure $p$, and velocity $\mbf{v}$ that is heated externally and cooled radiatively,
\begin{equation}\label{EOC}
\frac{\partial \rho}{\partial t}+\frac{\partial}{\partial x_{k}}(\rho v_{k})=0 \, ,
\end{equation}
\begin{equation} \label{EOM}
\frac{\partial \rho v_{i}}{\partial t}+\frac{\partial}{\partial x_{k}}
(\rho v_{i} v_{k}+p \delta_{ik})=0 \, ,
\end{equation}
\begin{equation} \label{EOE}
\frac{\partial E}{\partial t}+\frac{\partial}{\partial x_{k}}
\left\{ (E+p)v_{k}-\kappa \frac{\partial T}{\partial x_{k}}\right\}=-\rho \mathcal{L}(\rho,\,T) \, ,
\end{equation}
\begin{equation} \label{EOS}
p=\frac{R}{\mu} \rho T \, , 
\end{equation}
\begin{equation} \label{ST}
E=\frac{p}{\gamma -1}+\frac{\rho v_{l}v_{l}}{2} \, ,
\end{equation}
Here $\mathcal{L}$ is the net cooling function, $\kappa$ is the thermal conductivity, $\gamma$ is the ratio of specific-heats, $R$ is the gas constant, and $\mu$ is the mean molecular weight.
We adopt the following simplified cooling function
\begin{equation}\label{CF1}
\rho\,\mathcal{L} = n\,(-\Gamma +n\,\Lambda)\,, 
\end{equation}
where
\begin{eqnarray}\label{CF2}
\Gamma &=& 2\times 10^{-26} \,\,\mbox{ergs s}^{-1}\,,\\ \label{CF3}
\Lambda &=& 7.3\times 10^{-21}\,\exp\left( \frac{-118400}{T+1500} \right) + 
7.9\times 10^{-27}\,\exp\left( \frac{-92}{T} \right)\,\,\mbox{ergs cm}^{3}\,\mbox{s}^{-1}\,.
\end{eqnarray}
Figure \ref{f1} shows the equilibrium state of this cooling function ($\mathcal{L}(n,T)=0$) in the number density-pressure plane.
A two-phase structure is possible in the case that $10^{2.8}\la p/k_{B} \la 10^{4.1}$ K cm$^{-3}$ for this simplified choice of the cooling function.

In this section, we consider the steady state transition layer (or front) that connects the CNM and WNM in plane-parallel geometry, where the CNM and WNM are in a state of radiative equilibrium ($\mathcal{L}=0$).
From equations (\ref{EOC})-(\ref{EOE}), the state on either side of the front must satisfy the following conservation laws:
\begin{eqnarray}
\left[ \, \rho \, v_{x} \, \right ]=0 \, , \label{RH1} \\ 
\left[ \, p+\rho \, v_{x}^{2} \, \right ]=0 \, ,\label{RH2} \\
\left[ \, (E+p) \, v_{x} \, \right ]=Q \,, \label{RH3}
\end{eqnarray}
where brackets represent a difference across the front ($[F]=F(x=\infty)-F(x=-\infty)$) and
\begin{equation}\label{Q}
Q\equiv -\int \rho \mathcal{L}\,dx \,,
\end{equation}
is the net energy acquired as a result of external heating and radiative cooling.
Shchekinov \& Ib\'a\~nez (2001) give a detailed study of the solutions of equations (\ref{RH1})-(\ref{RH3}).
In the ISM, the following three types of solutions are important:
\begin{description}
\item[1.] Evaporation.--The CNM acquires energy through the front and evaporates to the WNM.
This is obtained by choosing the left- and right-hand states to be the CNM and WNM, respectively, with $Q>0$.
\item[2.] Condensation.--The WNM loses energy through the front and condenses to the CNM.
This is obtained by choosing the left- and right-hand states to be the WNM and CNM, respectively, with $Q<0$.
\item[3.] Saturation.--This is the static solution for which external heating and radiative cooling are balanced inside the front ($Q=0$).
\end{description} 

\subsection{Solutions Connecting Two Radiative Equilibrium States}

In this section, we calculate the structure of the fronts that connect the CNM and WNM.
The energy equation (\ref{EOE}) can be rewritten into the form
\begin{eqnarray}\label{EOE2}
\frac{\gamma}{\gamma -1}\,\frac{R}{\mu}\,\rho\,\frac{dT}{dt}-\frac{dp}{dt}=
\mbf{\nabla}\cdot \kappa \mbf{\nabla}T -\rho\,\mathcal{L}(\rho,T) \,,
\end{eqnarray}
where $d/dt=\partial/\partial t+\mbf{v}\cdot\mbf{\nabla}$.
We seek a one-dimensional solution that satisfies thermal equilibrium at infinity:
\begin{eqnarray} \label{TE}
\mathcal{L}(\rho_{1},T_{1})=0\,,\\
\mathcal{L}(\rho_{2},T_{2})=0\,,
\end{eqnarray}
where subscripts 1 and 2 denote the values at $x=-\infty$ and $x=\infty$, respectively.
Omitting the $y-,\,z-$ and time dependences of the variables, and integrating equations (\ref{EOC}) and (\ref{EOM}) with respect to $x$, we obtain the basic equations corresponding to the three conservation laws:
\begin{eqnarray}\label{SE1}
&& \rho \,v=\rho_{1}\,v_{1} \equiv j \, , \\ \label{SE2}
&& \rho\,v^{2}+p=\frac{j^{2}}{\rho_{1}}+p_{1}\equiv M \, , \\ \label{SE3}
&& \frac{\gamma}{\gamma -1}\,\frac{R}{\mu}\,j\,\frac{dT}{dx}-v\,\frac{dp}{dx}=
\frac{d}{dx}\left( \kappa \frac{d T}{d x} \right)-\rho\,\mathcal{L}(\rho,T) \,,
\end{eqnarray}
where $v$ is the $x$-component of the velocity.
From equations (\ref{SE1}),(\ref{SE2}), and (\ref{EOS}), the velocity, pressure, and density can be expressed in terms of the temperature, mass flux $j$ and total momentum $M$:
\begin{eqnarray} \label{Sv}
&&v = \frac{1}{2\,j}\Bigg( M-\sqrt{M^{2}-4\,\frac{R}{\mu}\,j^{2}\,T} \Bigg) \,, \\ \label{Sp}
&&p = M-jv=\frac{1}{2}\Bigg( M+\sqrt{M^{2}-4\,\frac{R}{\mu}\,j^{2}\,T} \Bigg) \,, \\ \label{Sr}
&&\rho = \frac{\mu}{R}\frac{P}{T}=\frac{\mu}{2RT}\Bigg( M+\sqrt{M^{2}-4\,\frac{R}{\mu}\,j^{2}\,T} \Bigg) \,.
\end{eqnarray}
The density $\rho_{1}$ and temperature $T_{1}$ at $x=-\infty$ are obtained from the solution of equation (\ref{TE}) if the pressure at $x=-\infty$ $(p_{1})$ is given.
Then the total momentum can be expressed as a function of the mass flux and given pressure at $x=-\infty$, $M=M(j,p_{1})$, and the boundary conditions
\begin{eqnarray}\label{BC1}
&& T(x=-\infty)=T_{1} \,,\\ \label{BC2}
&& \frac{dT}{dx} \bigg|_{x=\pm \infty} =0 \,,
\end{eqnarray}
are imposed on equation (\ref{SE3}).
Equation (\ref{SE3}) is a second-order ordinary differential equation with respect to $T$, whereas three boundary conditions (\ref{BC1}) and (\ref{BC2}) are imposed on it.
Therefore, it has a solution only when the mass flux $j$ is tuned as an eigenvalue of the solution, and $j$ is expressed as a function of $p_{1}$ (Shchekinov \& Ib\'a\~nez 2001).

In the case of a static front, that is, $j=v=0$, equation (\ref{SE3}) can be transformed as follows:
\begin{equation} \label{SAT}
\int^{T_{2}}_{T_{1}} \kappa\,\rho\,\mathcal{L}\,dT=\left[ 
\frac{1}{2}\,\kappa^{2}\left( \frac{dT}{dx} \right)^{2}
\right]^{x=+\infty}_{x=-\infty} =0\,
\end{equation}
(Zel'dovich \& Pikel'ner 1969, Penston \& Brown 1970).
Equation (\ref{SAT}) expresses the balance between cooling and heating inside a static front. The pressure that allows such a static structure is called the \textit{saturation pressure}.
Substituting equations (\ref{CF1})-(\ref{CF3}) into equation (\ref{SAT}) and solving, we obtain a saturation pressure $p_{s}/k_{B}=2612$ K cm$^{-3}$ for our choice of cooling function.
If $p_{1}$ is higher (lower) than this pressure, fluid elements that pass through the front experience a net cooling (heating) inside the front.
Thus, such a front becomes a condensation (evaporation) front.

\begin{figure}
\epsscale{.65}
\plotone{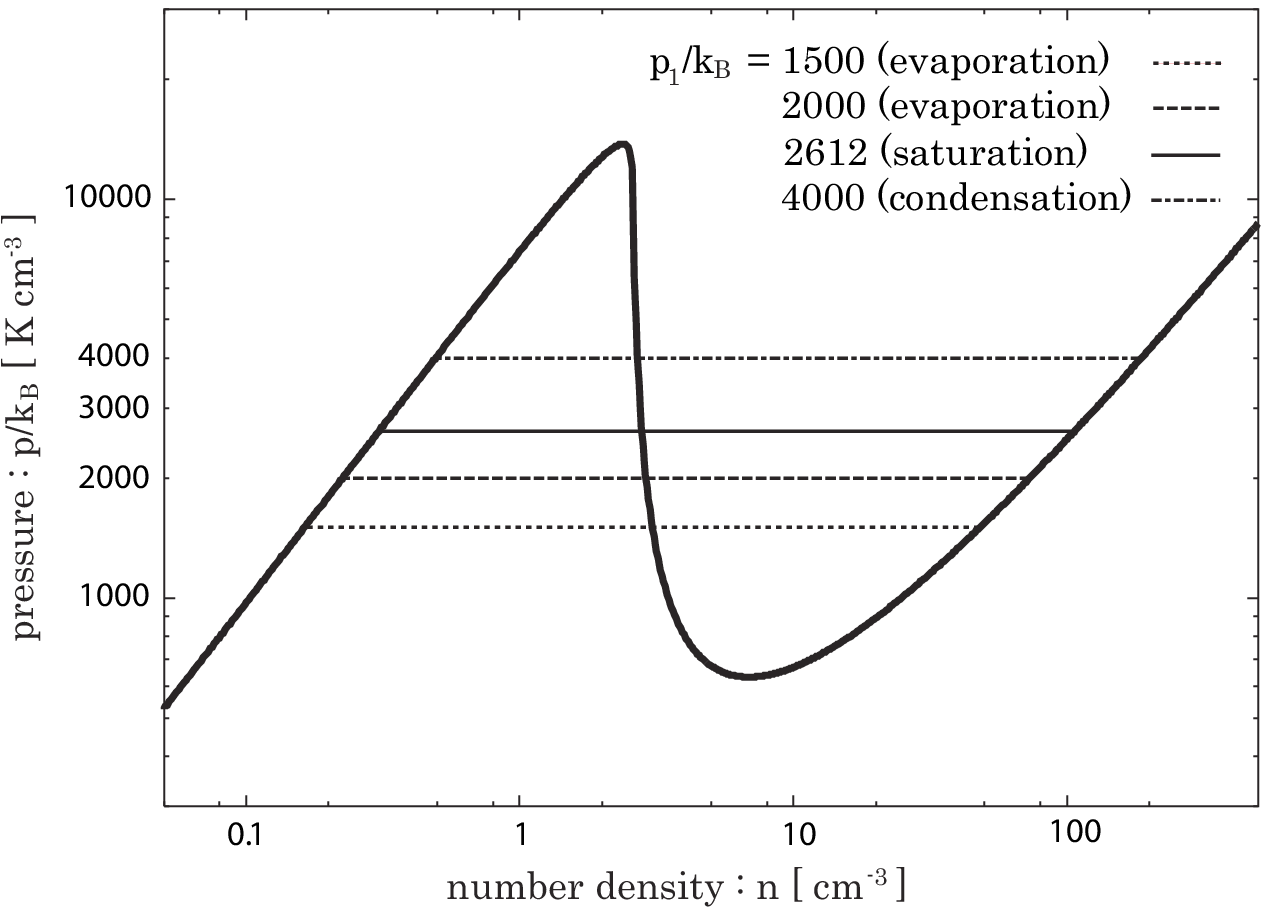}
\caption{
Equilibrium state with the cooling function given by eqs (\ref{CF1})-(\ref{CF3}) (thick solid line).
The thin solid line shows the steady solution of a saturation front ($p/k_{\rm B}=2612$ K cm$^{-3}$), the evaporation solutions for $p_{1}/k_{\rm B}=1500,$ and $2000$ K cm$^{-3}$ are plotted as dotted and dashed lines, respectively.
The condensation solution for $p_{1}/k_{\rm B}=4000$ K cm$^{-3}$ is plotted as a dot-dashed line.
\label{f1}}
\end{figure}

\begin{figure}
\epsscale{.65}
\plotone{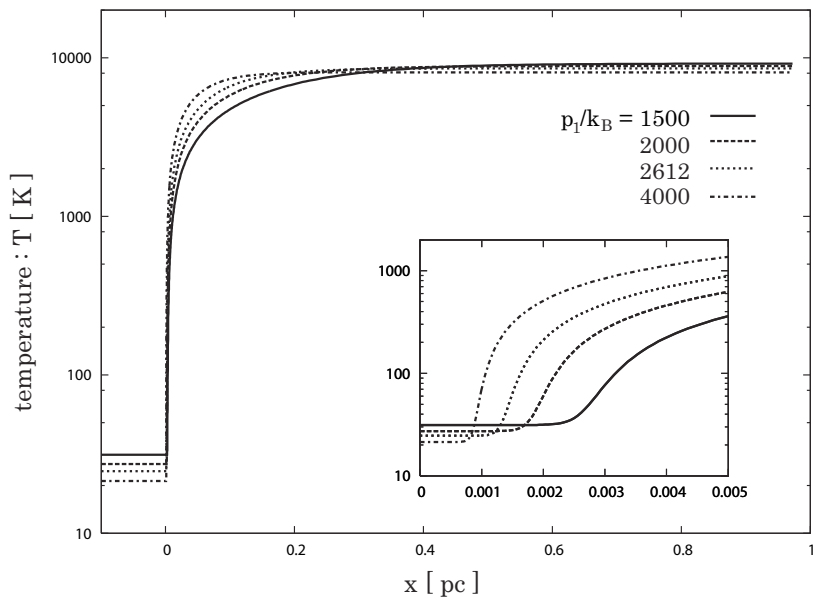}
\caption{
Temperature structure of the steady solutions.
The solid, dotted, dashed and dot-dashed lines show the solutions for $p_{1}/k_{\rm B}=1500$ (evaporation), $2000$ (evaporation), $2612$ (saturation), and $4000$ (condensation) K cm$^{-3}$, respectively.
\label{f2}}
\end{figure}

\begin{figure}
\epsscale{.65}
\plotone{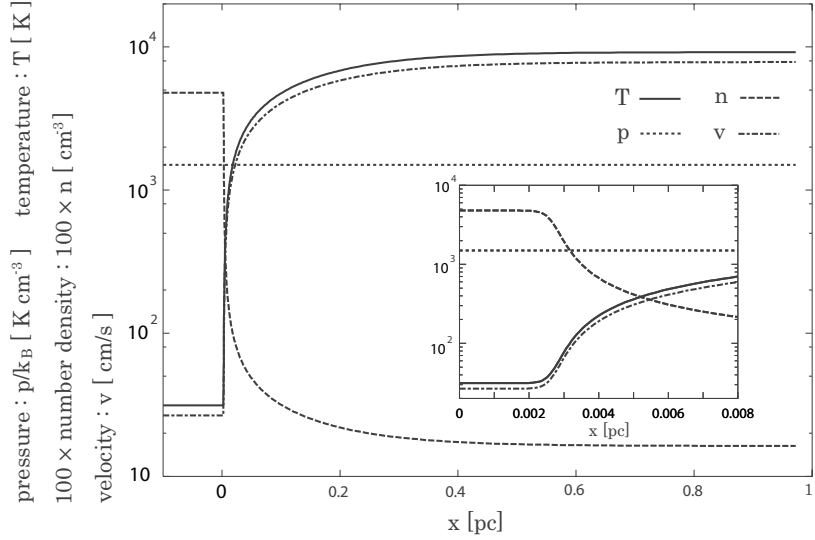}
\caption{
Temperature $T$ (solid line), pressure $p$ (dotted line), number density $n$ (dashed line) and velocity $v$ (dot-dashed line) structures of the steady evaporation solution for $p_{1}/k_{\rm B}=1500$ K cm$^{-3}$, where the number density has been multiplied by 100. 
\label{f3}}
\end{figure}

We numerically solve equations (\ref{SE1})-(\ref{SE3}) with the boundary conditions of equations (\ref{BC1}) and (\ref{BC2}).
We take the CNM to be state 1.
The dominant contribution to thermal conductivity is due to neutral atoms, so we use $\kappa=2.5\times 10^{3}\,T^{1/2}$ ergs s$^{-1}$K$^{-1}$cm$^{-1}$ (Parker 1953).
We plot the solutions in the number density-pressure diagram and in terms of temperature structure in Figures \ref{f1} and \ref{f2}, respectively, for values $p_{1}/k_{\rm B}$ of $1500, 2000, 2612,$ and $ 4000$ K cm$^{-3}$.
The structure of the temperature, pressure, number density, and velocity for $p_{1}/k_{\rm B}=1500$ K cm$^{-3}$ are plotted in Figure \ref{f3}.

Figures \ref{f1} and \ref{f3} show that the structure is almost isobaric.
That is due to the small Mach number of the flow (typically $\mathcal{M}\simeq 10^{-2}$).
From equation (\ref{SE1}) and (\ref{SE2}), the pressure difference between the CNM and WNM is
\begin{eqnarray}
\frac{p_{\rm c}-p_{\rm w}}{p_{\rm w}}&=&\frac{j\,(v_{\rm w}-v_{\rm c})}{p_{\rm w}} \\
 &\simeq&\gamma\,\left( \frac{v_{\rm w}}{c_{\rm w}} \right)^{2} = 10^{-4}\,.
\end{eqnarray}
where subscripts $\rm w$ and $\rm c$ denote the values in the WNM and CNM, respectively, $c=(\gamma\,R\,T/\mu)^{1/2}$ is the sound speed, and we use the fact that $v_{\rm w}\gg v_{\rm c}$. 

From equation (\ref{SE3}), the thickness of the front is determined by the Field length $l_{F}$,
\begin{equation}
\Delta x \sim \sqrt{\frac{\kappa T}{\rho \mathcal{L}}}=l_{F}\,
\end{equation}
(Begelman \& McKee 1990; Ferrara \& Shchekivov 1993),
which is a function of local variables and therefore takes different values in the CNM and WNM:
\begin{eqnarray}
&& l_{F,\rm c} = \sqrt{\kappa T_{\rm c}/n_{\rm c}^{2} \Lambda}\sim 10^{-2}\,\mbox{pc}\,,\\
&& l_{F,\rm w} = \sqrt{\kappa T_{\rm w}/n_{\rm w} \Gamma}\sim 10^{-1}\,\mbox{pc}\,,
\end{eqnarray}
This property can be seen in Figures \ref{f2} and \ref{f3}.
The total thickness is essentially determined by the Field length in the WNM, $l_{F,\rm w}$.

\begin{figure}
\epsscale{.60}
\plotone{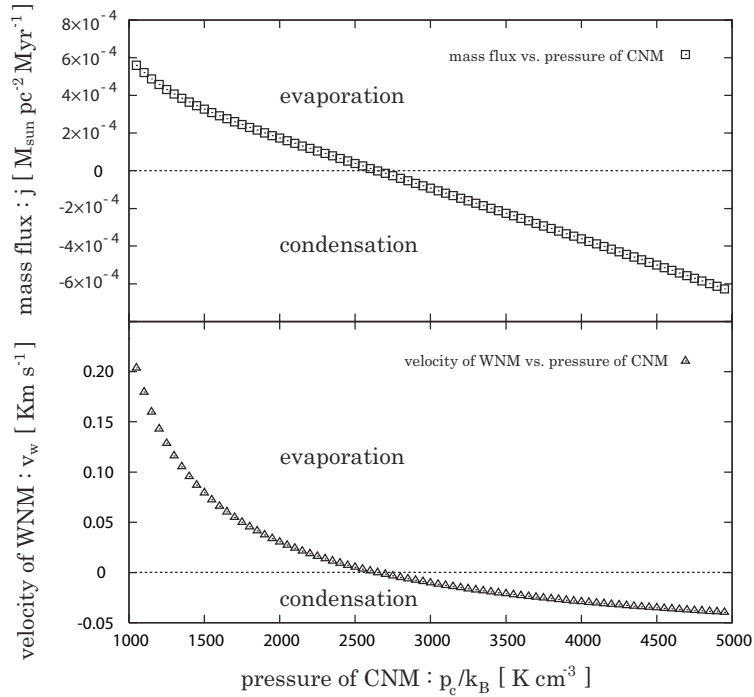}
\caption{
Mass flux $j$ (top) and velocity of the WNM $v_{\rm w}$ (bottom) vs. pressure of the CNM $p_{1}$.
Positive velocities correspond to evaporation.
\label{f4}}
\end{figure}

In Figure \ref{f4}, we plot the mass flux $j$ and the velocity of the WNM $v_{\rm w}$ as functions of $p_{1}$ in the top and bottom panels, respectively.
The typical evaporation flow speed is on the order of $0.1$ km s$^{-1}$ for this type of solution connecting two radiative equilibrium states.

\subsection{Solutions with Finite Spatial Extent}

\begin{figure}
\epsscale{.65}
\plotone{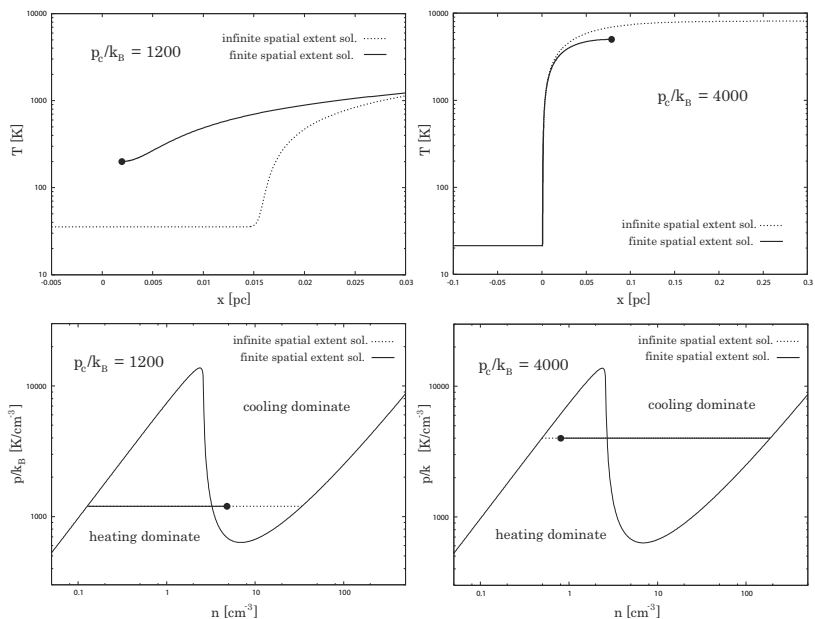}
\caption{
Temperature structure (top) and $n-p$ diagram (bottom) for the finite spatial extent solutions.
The left panels show the evaporation solution for $p/k_{\rm B}=1200$ K cm$^{-3}$, and the right panels show the condensation solution at which $p/k_{\rm B}=4000$ K cm$^{-3}$.
We also plot the infinite spatial extent solutions connecting the two radiative equilibrium states (dotted lines).
\label{f5}}
\end{figure}

We can obtain a larger evaporation flow speed (or smaller condensation flow speed), if we change the boundary condition in equation (\ref{BC1}) to a higher CNM temperature, without changing $p_{1}$, and impose the CMN boundary conditions at a finite distance from the front.
When such conditions are imposed, the CNM is no longer in a state of thermal equilibrium: $\mathcal{L}(\rho_{\rm c},T_{\rm c})\neq 0$.
The top and bottom left panels of Figure \ref{f5} show, respectively, the temperature structure and $n-p$ diagram for solutions of infinite and finite spatial extent with $p_{\rm c}/k_{\rm B}=1200$ K cm$^{-3}$.
The CNM temperature and evaporation flow speed are $T_{\rm c}=250 $K, and $v_{\rm w}=0.34$ km s$^{-1}$ for the finite spatial extent solution and $T_{\rm c}=35.5 $K and $v_{\rm w}=0.14$ km s$^{-1}$ for infinite one.
In both solutions, the WNM temperatures are the same.
From the bottom left panel of Figure \ref{f5}, one can see that the net heating of a fluid element that passes through the front is stronger for the finite spatial extent solution than for the infinite one, because the cooling becomes weaker.
Thus the evaporation flow speed becomes faster when the CNM temperature is larger than the thermal equilibrium temperature.

Of course, we can obtain a lower WNM temperature solution if we impose the boundary conditions for the WNM at a finite distance from the front.
In this case, the condensation flow speed becomes large (or the evaporation flow speed becomes small), because of the stronger net cooling (weaker heating) inside the front.
The top and bottom right panels of Figure \ref{f5} show, respectively, the temperature structure and $n-p$ diagram of the infinite and finite spatial extent solutions for $p_{\rm c}/k_{\rm B}=4000$ K cm$^{-3}$.
The WNM temperature and condensation flow speed are $T_{\rm w}=5000 $K, and $v_{\rm w}=-3.73\times 10^{-2}$ km s$^{-1}$ for the finite spatial extent solution and $T_{\rm w}=8115 $K, and $v_{\rm w}=-3.02\times 10^{-2}$ km s$^{-1}$ for the infinite one.
In both solutions, the CNM temperatures are the same.
These kinds of solutions are naturally realized in dynamical environments.

\begin{figure}
\epsscale{.65}
\plotone{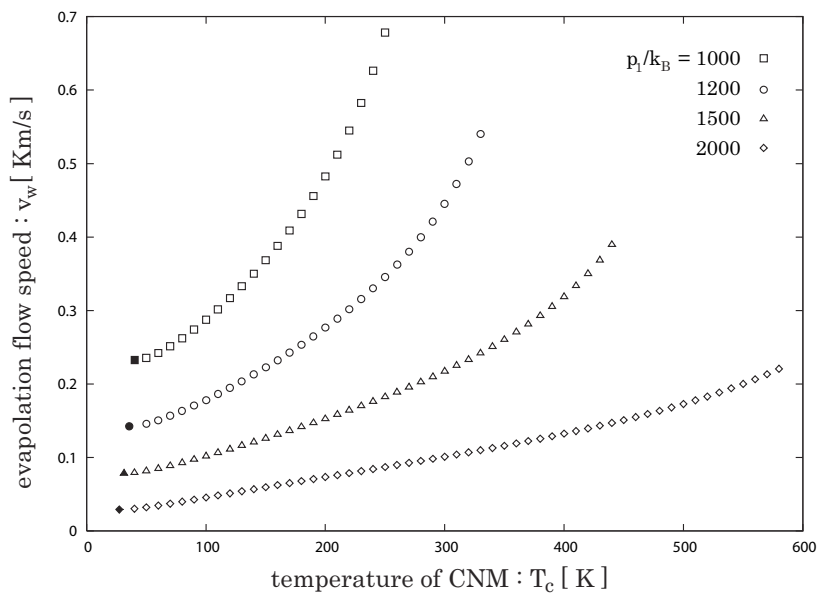}
\caption{
Evaporation flow speed $v_{\rm w}$ vs. temperature of the CNM in the case that $p_{\rm c}/k_{\rm B}=1000$ (squares), $1200$ (circles), $1500$ (triangles), and $2000$ (diamonds) K cm$^{-3}$.
Filled points correspond to the thermal equilibrium temperature of the CNM.
\label{f6}}
\end{figure}

We plot the evaporation flow speed $v_{\rm w}$ versus the temperature of the CNM in the case that $p_{1}/k_{B}=1000,\,1200,\,1500,$ and $2000$ K cm$^{-3}$ in Figure \ref{f6}.
The evaporation flow can reach speeds as large as 1 km s$^{-1}$.
In \S 3, we show that the growth rate of the DLI (eq. (\ref{GRDL})) is proportional to the evaporation flow speed $v_{\rm w}$.
Thus, this effect makes the DLI more effective.

\section{STABILITY OF PHASE TRANSITION LAYERS}

\subsection{Long-Wavelength Approximation}

In this section, we focus on perturbations on the scales much larger than the thickness of the front (the Field length), so we treat the transition layer as a discontinuous front.
The effect of the finite thickness of the transition layer is considered in the \S 3.2.

\begin{figure}
\epsscale{.60}
\plotone{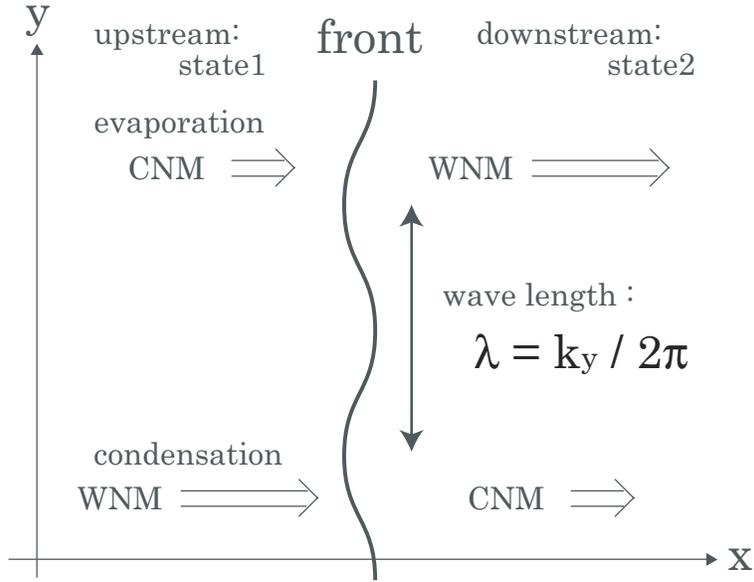}
\caption{
Schematic illustration of the perturbed front.
We denote the wavenumber of the perturbed position of the front as $k_{y}$.
We always set the left- and right-hand states as upstream and downstream, respectively.
\label{f7}}
\end{figure}

We assume that the discontinuous front is located in the plane $x=0$ and that the direction of the flow is parallel to the $x$-axis, and we choose a reference frame that moves together with the front.
The unperturbed state satisfies the jump conditions (\ref{RH1})-(\ref{RH3}).
As stated in the previous section, the jump conditions admit three types of solution, that is, evaporation, condensation, and saturation solutions.
In the following, we carry out a linear stability analysis of these fronts under a corrugational deformation.
The situation is illustrated schematically in Figure \ref{f7}.
We assume that perturbations of the fluid on both sides of the front are incompressible, because the flow speed is sub-sonic in either region.
The basic perturbed equations of the fluid are
\begin{eqnarray}\label{PLR1}
&& k_{x}\,\delta v_{x}+k_{y}\,\delta v_{y}=0 \, , \\ \label{PLR2}
&& \rho_{1,2}\,(\omega \, \delta v_{x}+k_{x}\,v_{1,2}\,\delta v_{x}) =-k_{x}\,\delta p \, , \\ \label{PLR3}
&& \rho_{1,2}\,(\omega \, \delta v_{y}+k_{x}\,v_{1,2}\,\delta v_{y}) =-k_{y}\,\delta p \, ,
\end{eqnarray}
where subscript 1 and 2 denote the left ($x<0$) and right ($x>0$) sides of the unperturbed state, respectively, and we assume that the perturbation is proportional to $e^{i\,(k_{x}\,x+k_{y}\,y-\omega \,t)}$.
From equations (\ref{PLR1})-(\ref{PLR3}), the perturbations in the pressure and the $x$-component of the velocity can be expressed in terms of the $y$-component velocity perturbation:
\begin{eqnarray} \label{PL1}
&&\delta p = -\frac{\rho_{1}\,(k_{x}^{(-)}\,v_{1}-\omega)}{k_{y}}\,\delta v_{y}^{(-)} \, , \\ \label{PL2}
&&\delta v_{x} = -\frac{k_{y}}{k_{x}^{(-)}}\,\delta v_{y}^{(-)} \, , \\ \label{PL3}
&&\delta v_{y} = \delta v_{y}^{(-)} \,, 
\end{eqnarray}
on the left side of the front, and
\begin{eqnarray} \label{PR1}
&&\delta p = -\frac{\rho_{2}\,(k_{x}^{(+)}\,v_{2}-\omega)}{k_{y}}\,\delta v_{y}^{(+)}
-\frac{\rho_{2}\,(k_{x}^{(s)}\,v_{2}-\omega)}{k_{y}}\,\delta v_{y}^{(s)} \, , \\ \label{PR2}
&&\delta v_{x} = -\frac{k_{y}}{k_{x}^{(+)}}\,\delta v_{y}^{(+)}-\frac{k_{y}}{k_{x}^{(s)}}\,\delta v_{y}^{(s)} \, , \\ \label{PR3}
&&\delta v_{y} = \delta v_{y}^{(+)}+\delta v_{y}^{(s)} \,,
\end{eqnarray}
on the right side, where $k_{x}^{(-)}=-i\,k_{y},\,k_{x}^{(+)}=i\,k_{y},$ and $k_{x}^{(s)}=\omega/v_{r}$.
Each of these modes damps as it propagates downstream (the $(+)$ and the $(s)$ modes) or upstream (the $(-)$ mode), where we have classified the $(s)$-mode as a downstream component because we seek an unstable solution (Im$[\omega]>0$).

Let us consider the matching conditions for the perturbations at the front.
The perturbation in the position of the front can be expressed as $\delta x_{f}\propto e^{i\,(k_{y}\,y-\omega\,t)}$, and thus the velocity perturbations normal and tangential to the front are given by
\begin{eqnarray}\label{PVH}
&&\delta v_{n}=\delta v_{x}+i\,\omega\,\delta x_{f}\,, \\ \label{PVT}
&&\delta v_{t}=\delta v_{y}+i\,k_{y}\,v\,\delta x_{f}\,.
\end{eqnarray}
The matching conditions of the front are perturbed version of mass and momentum flux conservation laws:
\begin{eqnarray}\label{PRH1}
[\rho\,\delta v_{n}]=0\,, \\ \label{PRH2}
[\delta p+2\,\rho\,v\,\delta v_{n}]=0 \,, \\ \label{PRH3}
[\rho\,v\,\delta v_{t}]=0\,.
\end{eqnarray}
We modify the equation (\ref{PRH1}) for the following physical reasons.
The discontinuous-front approximation can be justified when we consider a perturbation whose wavelength is longer than the thickness of the front; the structure of the front is hardly deformed under such a perturbation. 
Thus, the mass flux passing through the front will not changed, because it is determined by the structure of the front (eq. (\ref{Q}))
\begin{eqnarray}\label{PRH1b}
\rho_{1}\,\delta v_{n,1}=\rho_{2}\,\delta v_{n,2}=0\,.
\end{eqnarray}
Substituting equations (\ref{PL1})-(\ref{PVT}) into equations (\ref{PRH2})-(\ref{PRH1b}), we obtain the characteristic equation
\begin{equation}\label{DLM} M\,\mbf{\eta}=0\,, \end{equation}
where
\begin{equation}
\mbf{\eta}=(\delta x_{f},\delta v_{y}^{(-)},\delta v_{y}^{(+)},\delta v_{y}^{(s)})^{t}\,, 
\end{equation}
\begin{equation}
M=\left( \begin{array}{cccc}
\omega & -1 & 0 & 0 \\
\omega & 0 & 1 & i\,k_{y}v_{2}/\omega \\
0 & 1+(i\,\omega/r_{d} k_{y} v_{2}) & 1+(i\,\omega/k_{y} v_{2}) & 2\,i\,k_{y}v_{2}/\omega \\
i\,k_{y}(r_{d} -1)v_{2} & 1 & -1 & -1 \\
\end{array} \right) \,,
\end{equation}
and $r_{d} =\rho_{2}/\rho_{1}=v_{1}/v_{2}$ is the compression ratio between regions 1 and 2.
Taking the determinant of $M$, we obtain the dispersion relation for the front,
\begin{equation}\label{GRDL}
-i\,\omega = \frac{-r_{d}+r_{d}\sqrt{1-r_{d}+r_{d}^{-1}}}{1+r_{d}} \,v_{2}\,k_{y}.
\end{equation}
In the case of evaporation, $r_{d}=\rho_{\rm w}/\rho_{\rm c}<1$, and in case of the condensation, $r_{d}=\rho_{\rm c}/\rho_{\rm w}>1$.
Thus, the evaporation front is unstable and the condensation front is stable. 
Note that this is a similar to the Darrieus-Landau instability in combustion fronts (see, e.g., Landau \& Lifshitz 1987).
The similarity of the evaporation and the combustion fronts can be understood if one imagines the CNM as cold fuel, the WNM as hot exhaust, and $Q$ as the energy released by combustion.  

From equation (\ref{GRDL}), the growth rate of the instability is proportional to the wavenumber of the corrugation $k_{y}$.
Therefore, the most unstable scale is infinitesimal, and we cannot estimate the physical scale of the instability or its growth time.
This due to our neglect of the thickness of the front. 

To obtain a more realistic dispersion relation, Aranson et al. (1995) considered the effect of the curvature of the front (see, Graham \& Langer 1973; Aranson et al. 1995; Nagashima et al. 2005) and changed the right-hand side of equation (\ref{PRH1b}) to $a\,k_{y}$, where $a$ is a mass flux variance due to the curvature effect.
This stabilizes the front for the short wavelength modes, and is also known as the ``Markstein effect" in the case of a combustion front (Markstein 1964; Zel'dovich et al. 1985).

However, the most reliable way to determine realistic dispersion relation is to consider the thickness (characteristic scale) of the front in the analysis.
By doing so, we can account for the effects of thermal conduction and thermal stability inside the front that are lost in the discontinuous-front approximation.
In the following section, we therefore study the effect of the thickness of the front.

\subsection{Short-Wavelength Approximation}

In this section, we study the linear stability of the transition layer obtained in \S 2.1.
However, it is difficult to carry out the analysis without approximations.
Thus, we consider a perturbation whose wavelength is much shorter than the acoustic scale $l_{a}$, which is defined as $c_{s}\,t_{\rm c}$, where $c_{s}$ is the sound speed and $t_{\rm c}$ is the characteristic cooling (or heating) timescale.
Normally, the acoustic scale is longer than the Field length: $l_{a,\rm w}\sim 10$ pc in the WNM and $l_{a,\rm c}\sim 0.1$ pc in the CNM.
For such  small-scale modes, pressure balance sets in rapidly, and perturbations of the thermodynamic variables (temperature and density) are important.
On the other hand, perturbations of the fluid dynamical variables (pressure and velocity) are important, for the large-scale modes considered in \S 3.1.

We assume an isobaric perturbation of the front.
The basic equations are perturbed version of equations (\ref{EOE2}) and (\ref{EOS}) for the themodynamic variables:
\begin{eqnarray}
&& \frac{\gamma}{\gamma-1}\frac{R}{\mu}\left\{ \rho_{0} \left( n\,\delta T+v_{0}\,D\,\delta T\right)+v_{0}\,\left(D\,T_{0}\right)\,\delta \rho_{0} \right\} = \nonumber \\ \label{PEOE}
&& \ \ \ \  \kappa(T_{0}) \left( (D^{2}-k_{y}^{2})+\frac{(D\,T_{0})}{T_{0}}D+\frac{(D^{2}\,T_{0})}{2\,T_{0}}
-\frac{(D\,T_{0})^{2}}{4\,T_{0}^{2}} \right)\delta T
-L_{T}\,\delta T-L_{\rho}\,\delta \rho\,, \\ \label{PEOS} 
&& \frac{\delta T}{T_{0}}+\frac{\delta \rho}{\rho_{0}}=0\,,
\end{eqnarray}
where a subscript zero denotes an unperturbed variable, $D=d/dx$ is a differential operator, and we write $L_{T}\equiv \partial\,(\rho\,\mathcal{L})/\partial\,T$ and $L_{\rho}\equiv \partial\,(\rho\,\mathcal{L})/\partial\,\rho$.
We assume that a perturbed physical variable $f$ takes the form
\begin{equation}
f(\mbf{r},t)=f_{0}(x)+\delta f(x)\,\exp(i\,k_{y}\,y+n\,t)\,.
\end{equation}

Since the unperturbed state is spatially uniform at infinity, we can obtain asymptotic solutions by assuming solutions proportional to $e^{\chi\,x}$.
From equations (\ref{PEOE}) and (\ref{PEOS}), $\chi$ is determined by the following polynomial:
\begin{equation}\label{SWL1}
\kappa(T_{\rm c,\rm w})\,(\chi^{2}-k_{y}^{2})-\rho_{\rm c,\rm w}\,\frac{\gamma}{\gamma-1}\frac{R}{\mu}(n+v_{x\,\rm c,\rm w}\,\chi)-L_{T}|_{\rm c,\rm w}+\frac{\rho_{\rm c,\rm w}}{T_{\rm c,\rm w}}L_{\rho}|_{\rm c,\rm w}=0\,.
\end{equation}
We seek a localized solution around the front that vanishes at infinity.
Therefore, the asymptotic boundary conditions are
\begin{eqnarray}\label{SWL2}
\delta T(x)&=&C_{1}\,e^{\chi_{\rm c}\,x}\,, \\ \label{SWL3}
D\delta T(x)&=&C_{1}\,\chi_{\rm c}\,e^{\chi_{\rm c}\,x}\,,
\end{eqnarray}
far from the front in the CNM, and
\begin{eqnarray}\label{SWL4}
\delta T(x)&=&C_{2}\,e^{\chi_{\rm w}\,x}\,, \\ \label{SWL5}
D\delta T(x)&=&C_{2}\,\chi_{\rm w}\,e^{\chi_{\rm w}\,x}\,,
\end{eqnarray}
far from the front in the WNM, where $C_{1}$ and $C_{2}$ are constant, and Re$[\chi_{\rm c}]>0$ and Re$[ \chi_{\rm w}]<0$.
To avoid the freedom to translate the amplitude of the perturbation, we fix $C_{1}$.
A solution can be obtained if $k_{y}$ is given and $n$ and $C_{2}$ are eigenvalues of equations (\ref{PEOE}) and (\ref{PEOS}) with the above boundary conditions.

\begin{figure}
\epsscale{.65}
\plotone{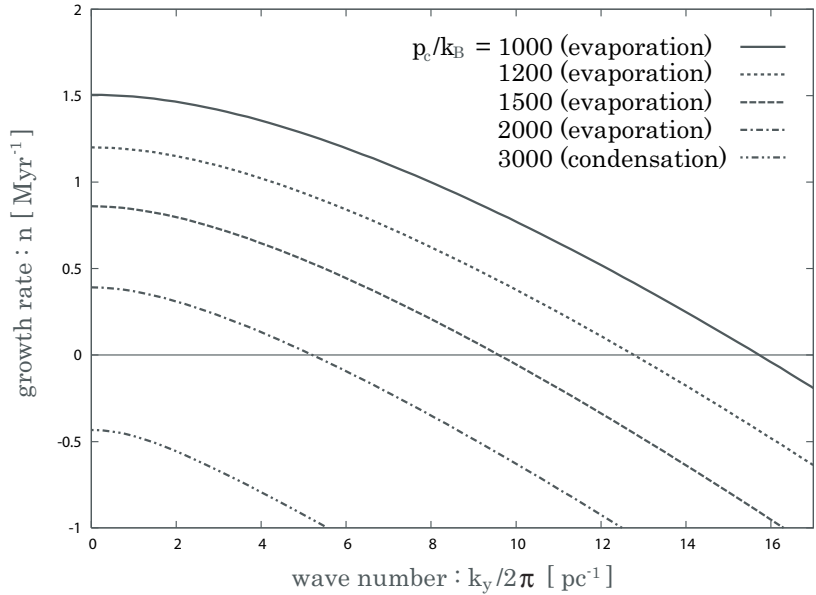}
\caption{
Dispersion relations for the isobaric perturbation.
The solid, dotted, dashed, dot-dashed, and dot-dot-dashed lines correspond to the unperturbed states of $p_{\rm c}/k_{\rm B}=1000$ (evaporation), $1200$ (evaporation), $1500$ (evaporation), $2000$ (evaporation), and $3000$ (condensation) K cm$^{-3}$, respectively.
\label{f8}}
\end{figure}

This eigenvalue problem has an infinite number of solutions (eigenmodes).
In the analysis, we are not interested in overstable modes.
Therefore, we search for a solution whose growth rate $n$ is real.
We numerically solve the eigenvalue problem for a given wavenumber.
The dispersion relations of the most unstable eigenmodes are plotted in Figure \ref{f8} for $p_{\rm c}/k_{B}=1500,\,2000\,,3000\,$ K cm$^{-3}$.
In the case of evaporation, there is a range of unstable wavenumbers, whereas all values are stable in the case of condensation.
In all cases, the eigenfunctions do not have a node; that is, the solutions are fundamental modes.
The timescale of the instability is roughly on the order of the cooling timescale of the WNM $t_{\rm c, \rm w}\sim 1$ Myr or shorter.
The instability is stabilized at a scale approximately equal to the thickness of the front, $l_{F,\rm w}\sim 0.1$ pc, as a result of thermal conduction.
In the isobaric approximation, results for wavelengths comparable to or longer than the acoustic scale are not valid.
We expect that the growth rate would be suppressed at longer wavelengths as in equation (\ref{GRDL}), if we were to not use the isobaric approximation.

\section{SUMMARY AND DISCUSSION}
We have performed a linear stability analysis of the steady transition layer that connects the CNM and WNM in both the long- and short-wavelength limits.
Our results show that the evaporation front that forms when the surrounding pressure is lower than the saturation pressure is unstable under corrugational deformations, whereas condensation fronts seem to be stable.

The result of the long-wavelength (discontinuous front) approximation, equation (\ref{GRDL}), shows that the growth timescale of the instability $t_{\rm grow}$ is inversely proportional to the speed of the evaporation flow $v_{\rm w}$ and the corrugation wavenumber $k_{y}$:
\begin{equation}\label{GR}
t_{\rm grow}\equiv n^{-1}= 0.3\,\left( \frac{v_{\rm w}}{0.5\mbox{km/s}} \right)^{-1}\left( \frac{\lambda}{0.1\mbox{pc}} \right)\,\mbox{Myr}\,,
\end{equation}
where $\lambda=2\,\pi/k_{y}$ is the wavelength of the corrugation.
A typical density contrast in the ISM is about a hundred, $r_{d}=0.01$.
The results of the short-wavelength (isobaric) approximation show that the instability stabilizes at the scale of the thickness of the front, $\sim 0.1$ pc, which is essentially determined by the Field length of the WNM.

\begin{figure}
\epsscale{.65}
\plotone{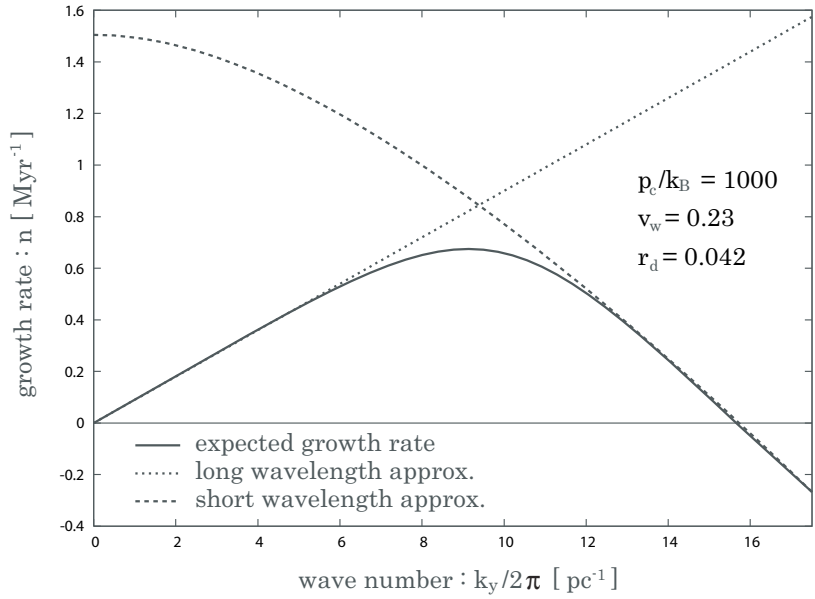}
\caption{
Dispersion relations of both the long- (dotted line) and short- (dashed line) wavelength approximations for $p_{\rm c}/k_{\rm B}=1000$ K cm$^{-3}$.
The unperturbed flow speed is $v_{\rm w}=0.23$ km s$^{-1}$, and the density contrast of the CNM and WNM is $r_{d}=0.042$.
The solid line is a schematic dispersion relation that is expected to be obtained from the analysis without approximations.
\label{f9}}
\end{figure}

We plot the dispersion relation for both the long- and short-wavelength approximations at a value of $p_{\rm c}/k_{B}=1000$ K cm$^{-3}$ in Figure \ref{f9}.
The dotted and dashed lines are, respectively, the dispersion relations in the long- and short-wavelength approximations.
The solid line is a schematic dispersion relation expected to be obtained from an analysis without approximations.
From this figure, we expect the most unstable scale to be about twice the stabilized scale.
For such a scale, the growth timescale of the instability is comparable to the cooling timescale of the WNM ($\sim 1$ Myr).
The growth timescale can be shorter in a dynamical environment because of the effect of finite spatial extent as studied in \S 2.2.
The evaporation flow speed becomes as large as $1$ km s$^{-1}$.
For this flow speed, from equation (\ref{GR}), the growth timescale can be as small as the cooling time-scale of the CNM ($\sim 0.1-0.01$ Myr).
Therefore, the instability is expected to grow within a dynamical timescale of the ISM ($\sim 1-10$ Myr) and may drive a turbulent two-phase medium in the nonlinear stage, as in the case of a combustion front.

Koyama \& Inutsuka (2002) and Audit \& Hennebelle (2005) have reported on the two-phase turbulence in numerical simulations of the thermally bistable medium, for which the velocity dispersion is on the order of the observed velocity dispersions in the diffuse ISM.
They propose that turbulence in the ISM can be driven in a thermally unstable gas constantly supplied by shock propagation (Koyama \& Inutsuka 2002) or converging flows (Audit \& Hennebelle 2005; see also Heitsch et al. 2005; V\'azquez-Semadeni et al. 2006).  
These shocks or converging flows are ubiquitously generated, for example, by supernovae, and correspond to energy input on large scales.
In contrast, our analysis of the stability of the transition layer shows another aspect of the dynamics of the multi phase medium that may be created by the thermal instability: the instability of the evaporating front can generate fluctuating motions even without external mechanical forcing.  
Note, however, that the maximum growth rate of the instability depends sensitively on the velocity of the evaporating flow.  
A further detailed study of the nonlinear behavior of the instability will be presented in our next paper.

\acknowledgments
We thank Masahiro Nagashima for useful discussions.
This work is supported by the Grant-in-Aid for the 21st Century COE "Center for Diversity and Universality in Physics" from the Ministry of Education, Culture, Sports, Science and Technology (MEXT) of Japan.
SI is supported by the Grant-in-Aid (No.15740118, 16077202, 18540238) from MEXT of Japan.
HK is supported by the 21st Century COE Program of Origin and Evolution of Planetary Systems in MEXT of Japan.

\clearpage

\clearpage

\end{document}